\documentclass[]{article}

\usepackage{amsthm} 
\usepackage{setspace}
\usepackage{bm} 
\usepackage{mathtools}
\usepackage{amsfonts}
\usepackage[authoryear]{natbib} 
\usepackage[letterpaper, margin=1in]{geometry} 
\usepackage{algorithm2e} 
\usepackage{float}
\allowdisplaybreaks 
\usepackage{graphicx} 
\usepackage{subcaption}
\usepackage{algorithm2e}

\newcommand{\Real}{\mathbb{R}}

\newcommand{\Tra}{^{\sf T}} 
\newcommand{\Inv}{^{-1}} 
 
\newcommand{\V}[1]{{\bm{\mathbf{\MakeLowercase{#1}}}}} 
\newcommand{\Vhat}[1]{{\bm{\hat \mathbf{\MakeLowercase{#1}}}}} 
\newcommand{\M}[1]{{\bm{\mathbf{\MakeUppercase{#1}}}}} 
\newcommand{\Mtilde}[1]{{\bm{\tilde \mathbf{\MakeUppercase{#1}}}}} 

\def\A{{\bf A}}

\def\C{{\bf C}} 
\def\c{{\bf c}} 

\def\D{{\bf D}} 
\def\d{{\bf d}}

\def\e{{\bf e}}

\def\F{{\bf F}}

\def\I{{\bf I}}

\def\U{{\bf U}}
\def\u{{\bf u}}

\def\w{{\bf w}}

\def\X{{\bf X}} 
\def\x{{\bf x}} 
\def\Xtilde{\Mtilde{X}}

\def\y{{\bf y}}

\def \vbeta{{\V{\beta}}}
\def \vpi{{\V{\pi}}}
\def \vmu{{\V{\mu}}}

\def \mlambda{{\M{\Lambda}}}

\def \vtheta{{\V{\theta}}}

\newcommand{\normal}{\mathcal{N}} 

\newcommand{\argmin}[1]{\underset{#1}{\text{ arg min }} }
\newcommand{\simiid}{\overset{iid}{\sim}}

\title{ \textbf{Dimension Reduction via Supervised Clustering of Regression
Coefficients: A Review} }
\author{Suchit Mehrotra \\ Department of Statistics \\ North Carolina State
University}

\date{}

\begin{document}

\maketitle

\doublespacing

\begin{abstract}
    The development and use of dimension reduction methods is prevalent in
    modern statistical literature.  This paper reviews a class of dimension
    reduction techniques which aim to simultaneously select relevant predictors
    and find clusters within them which share a common effect on the response.
    Such methods have been shown to have superior performance relative to OLS
    estimates and the lasso \citep{tibs1996lasso} especially when
    multicollinearity in the predictors is present. Their applications, which
    include genetics, epidemiology, and fMRI studies, are also discussed.
\end{abstract}

\section{Introduction}

Recent technological advances in computing power and data storage have
simplified the collection of vast quantities of data. Modern scientific
questions often involve statistical analyses on datasets with a large number of
predictors, many of which have no meaningful effect on the response. In
numerous situations, an analysis is further complicated by multicollinearity in
predictors. An example are microarray studies where interest lies in
understanding the relationship between a health outcome and gene expressions;
gene expression levels are often highly correlated and the number of genes is
substantially larger than the number of samples \citep{segal2003regression}. 

In a normal linear model, it is well know that ordinary least squares (OLS) is
not appropriate in high dimensional situations \citep{monahan2008primer}.
First, if the number of predictors is greater than the number of samples, a
unique solution to the OLS optimization problem does not exist. Additionally,
if a number of predictors are highly correlated, like in the microarray
example, inference in the model is unreliable \citep{faraway2016linear}. In
such situations, it may be reasonable to assume that highly correlated
predictors share the same effect on the response, if the correlation is due to
the same underlying factor. We can thus reduce the dimension of the problem by
summing or averaging groups of highly correlated columns and fitting a linear
model with the new set of predictors. 

This paper reviews methods which make a a slight generalization of this
assumption. The methods discussed assume that there exist groups of predictors
which have the same effect on the response, but these predictors need not be
correlated. Consequently, if we can identify groups with similar effects, we
can reduce the dimension of the problem and produce a model with parsimony and
superior predictive ability. In essence, this reduces to finding clusters in
the regression coefficients in a supervised manner.  We focus our attention on
the normal linear model, but the reviewed methods can be easily extended to
other types of responses. For the rest of this paper let $\y$ be a $N \times 1$
vector of responses, $\X$ a $N \times P$ matrix of covariates, $\vbeta$ a $P
\times 1$ vector of fixed effects, $\e$ a $N \times 1$ vector of errors, and
\begin{align} \label{eqn:lin-mod} \y = \X \vbeta + \e, \ \e \sim \normal(\V{0},
\sigma^2\I) \end{align} Additionally, assume that the columns of $\X$ are
standardized to mean zero and standard deviation one and $\y$ is centered with
mean zero. 

Many methods have been developed to address the problems posed by OLS in high
dimensional settings, with two common approaches being ridge regression and the
lasso. Ridge regression \citep{hoerl1970ridge} minimizes the residual sum of
squares with a restriction on the $\ell_2$ norm of the parameter vector
$\vbeta$.  This penalty has the effect of shrinking the elements of $\vbeta$ to
zero, but it does not produce a parsimonious model.  The lasso
\citep{tibs1996lasso} imposes an $\ell_1$ penalty on $\vbeta$ which encourages
variable selection by shrinking small effects to zero. Because of its variable
selection property, the lasso has become the first model of choice in many
applications.

However, there exist two common situations in which the lasso fails
\citep{zou2005enet}. If the predictor matrix $\X$ has a group of highly
correlated columns, the lasso arbitrarily selects only one from the group and
sets the other coefficients to zero. Additionally, if $P > N$, the lasso only
selects a maximum of $N$ variables. These issues are especially relevant in the
microarray example, where genes are not only correlated but many more genes
than samples might have an effect on the response. To alleviate these issues
\citet{zou2005enet} propose the elastic net, combining the $\ell_1$ penalty of
the lasso with an $\ell_2$ penalty, which estimates similar regression
coefficients for groups of correlated variables. 

While the elastic net can be effective in practice, it does not directly aim to
find clusters in $\vbeta$ and many methods have since been developed to improve
upon its performance.  These extensions aim to solve the supervised clustering
problem using three major approaches: two-step methods, discussed in Section
\ref{sec:two_step}, penalized regression methods, discussed in Section
\ref{sec:pen_methods}, and mixture models, discussed in Section
\ref{sec:mix_models}.

\section{Two-step approaches} \label{sec:two_step}

Much of the early work in supervised clustering of regression coefficients
focused on two-step procedures with distinct clustering and model fitting
steps. Since the correct grouping of predictors with the same effects is not
known \textit{a priori}, researchers made the assumption that correlated
predictors have correlated effects.  Consequently, the work discussed in this
section attempts to solve the problem by first finding a clustering of the
predictors, averaging the predictors in the same cluster to create a
``super-predictor", and using them to fit a linear model. 

\citet{hastie2001harvest} and \citet{park2006averaged} perform a two-step
procedure which combines hierarchical clustering with variable selection
methods, such as stepwise selection and the lasso. Algorithm \ref{alg:two_step}
details the method used by \citet{park2006averaged}. Two parameters in the
algorithm need to be estimated from the data using cross validation: the level
at which to cut the tree, $p$, and the tuning parameter for the lasso,
$\lambda$.

\RestyleAlgo{boxruled}
\LinesNumbered
\begin{algorithm}[ht]
    \caption{Hierarchical Clustering and Averaging for Regression
    \citep{park2006averaged}} \label{alg:two_step}
    Generate a dendogram by hierarchically clustering the predictors, creating
    a hierarchy of $P$ possible clusterings.  \\
    \For{$p = 1$ to $P$} {
        Create a new response matrix, $\Xtilde_p$, by averaging all the
        clusters at level $p$. \\ 
        Regress $\y$ on $\Xtilde_p$ using the lasso, with tuning parameter
        $\lambda$, to get estimate $\Vhat{\beta}_{p, \lambda}$
    }  
    Choose $p$ and $\lambda$ using cross-validation
\end{algorithm}

Another two-step algorithmic approach is given by \citet{jorn03simul}, who use
an iterative method to simultaneously cluster genes and classify a response of
interest.  The first part of their model is a clustering step which uses a
Gaussian mixture model and the second averages columns in a cluster to create
new predictors for classifying the response. \citet{dettling2004gene} propose a
similar method with an algorithm which iterates between clustering the
predictors into groups and fitting a logistic regression model. A more recent
two-step approach is proposed by \citet{buhlmann2013correlated}, who first
cluster the columns of $\X$ and then use these groups for parameter estimation
via the group lasso \citep{yuan2006group}.

There are multiple drawbacks to the procedures described above. First,
averaging correlated predictors is only a proxy for the supervised clustering
of regression coefficients, not a direct solution to the problem. If two
correlated predictors have different effects, it is easy to see how the
two-step approach can lead to poor estimates. Second, parameter estimation for
these methods requires solving a non-convex optimization problem. Hierarchical
clustering, used by \citet{park2006averaged}, is performed using heuristic
based agglomerative algorithms and always provides a clustering of the data
even if none exists \citep{murphy2012machine}. Third, two-step procedures do
not lend themselves to clear uncertainty quantification. Even if we replace the
lasso in Algorithm \ref{alg:two_step} with a linear model, standard error
estimates given by linear model theory would not account for the uncertainty in
the clustering of $\X$. 

The penalization based approaches considered in the next section effectively
resolve the first two issues. They impose penalties which encourage
coefficients to shrink to one another and allow for estimation of $\vbeta$ by
solving a convex optimization problem. 

\section{Penalization Based Methods} \label{sec:pen_methods}

The primary approach for dimension reduction in the frequentist literature is
to induce sparsity in $\vbeta$ via penalized regression. An estimate for the
coefficients is found by solving an optimization problem which minimizes the
residual sum of squares along with a penalty on the coefficients.  Generally,
the loss function can be written in the form of Equation \ref{eqn:loss+penalty}
where $\Omega(\vbeta)$ is a penalty term used to impose structure on $\vbeta$. 
\begin{align} \label{eqn:loss+penalty}
    \mathcal{L}(\vbeta, \lambda) = \frac{1}{2} ||\y - \X \vbeta||^2_2 + 
        \lambda \Omega(\vbeta)
\end{align}

The ideal penalty to induce sparsity in this context is the $\ell_0$ norm,
$\Omega(\vbeta) = ||\vbeta||_0$, which counts the number of non-zero entries in
$\vbeta$. However, this penalty leads to a NP-Hard optimization problem and is
computationally infeasible even for relatively small data sets
\citep{she2010sparse}.  A popular convex relaxation to the $\ell_0$ norm is the
lasso which uses $\Omega(\vbeta) = ||\vbeta||_1 = \sum_{j =1}^P |\beta_j|$
\citep{tibs1996lasso}.

The lasso's $\ell_1$ penalty can be can be modified to impose general
structural constraints which encompass a wide variety of problems. The modified
optimization problem is called the generalized lasso
\citep{tibshirani2011genlasso} and its solution is given by
\begin{align} \label{eqn:gen_lasso}
    \Vhat{\beta} = \argmin{\vbeta \in \Real^{P}} 
        \frac{1}{2} ||\y - \X \vbeta||^2_2 + 
        \lambda ||\D \vbeta||_1
\end{align}
where $\D \in \Real^{M \times P}$ is a prespecified penalty matrix. We will see
below that most of the penalization methods discussed can be cast into the
generalized lasso framework. This allows for efficient computation of parameter
estimates via the use of a path algorithm developed by
\citet{tibshirani2011genlasso}. 

Before we list specific methods which perform supervised clustering of
regression coefficients, it is useful to discuss the general framework for
finding clusters via convex penalized regression.  This will motivate the types
of penalties used in the regression methods below. 

\subsection{Convex Clustering}

Solutions to common clustering algorithms such as $k$-means and hierarchical
clustering are found by optimizing a non-convex function. This not only makes
the problem computationally intensive, but their solutions can also suffer from
instability. Recent advances \citep{lindsten2011just, hocking2011clusterpath}
aim to remedy these issues by using a convex relaxation of the clustering
problem. Clusters in the data can be found via penalized regression for a
particular choice of the tuning parameter. Given $N$ points, $\x_1, \dots,
\x_N$ in $\Real^m$, they optimize
\begin{align} \label{eqn:cvx_clust}
    \mathcal{L}(\U, \lambda, \w) & = 
        \frac{1}{2} \sum_{i = 1}^N ||\x_i - \u_i||^2_2 + 
        \lambda \sum_{i < j} w_{ij} ||\u_i - \u_j||_p
\end{align}
where $\lambda > 0$ is a tuning parameter, $w_{ij}$ is a non-negative weight,
$||\cdot||_p$ is an arbitrary norm, and $\u_i$ is the cluster center for point
$\x_i$.  The penalty on the differences $(\u_i - \u_j)$ forces the cluster
centers towards one another and, as $\lambda \to \infty$, $\u_i = \u_j$ for all
$i \neq j$. A clustering result similar to hierarchical clustering can also be
generated by solving Equation \ref{eqn:cvx_clust} at different values of
$\lambda$; as $\lambda$ goes from zero to $\infty$ the number of clusters goes
from $N$ to 1. Additionally, for given $\lambda$ and $\w$, Equation
\ref{eqn:cvx_clust} has a unique global minimizer and can be solved with fast
iterative algorithms developed by \citet{chi2015splitting}.

Equation \ref{eqn:cvx_clust} is useful in seeing how penalizing differences in
the elements of $\vbeta$ can promote clustering of regression coefficients.
The first use of such penalties was the fused lasso
\citep{tibshirani2005sparsity} which penalizes coefficients of neighbouring
covariates. Parameter estimates for the fused lasso are given by:
\begin{align} \label{eqn:fused_lasso}
    \Vhat{\beta} = \argmin{\vbeta \in \Real^P}
        \frac{1}{2} ||\y - \X \vbeta||_2^2 + 
        \lambda_1 ||\vbeta||_1 + 
        \lambda_2 \sum_{j = 2}^P |\beta_j - \beta_{j - 1}|
\end{align}
This approach incorporates variable selection and clustering into the same
optimization problem. The first term in the penalty  encourages sparsity in the
coefficients, while the second encourages sparsity in successive differences.
Like with convex clustering, as $\lambda_2 \to \infty$ successive differences
in $\vbeta$ go to zero and $\beta_i = \beta_j$ for all $i \neq j$.

It can easily be seen that the fused lasso problem given by Equation
\ref{eqn:fused_lasso} can be can be cast as a generalized lasso, albeit with
two tuning parameters. For example, if $\vbeta \in \Real^3$ the $\D$ which
makes Equation \ref{eqn:fused_lasso} equivalent to Equation \ref{eqn:gen_lasso}
is
\begin{align} \label{eqn:fused_D}
    \D_{\lambda_2} = 
    \begin{bmatrix}
        \I \\
        \lambda_2 \F
    \end{bmatrix} 
    \text{ where } 
    \F = 
    \begin{bmatrix}
        -1 & 1 & 0 \\
        0 & -1 & 1 
    \end{bmatrix}
\end{align}
Consequently, the fused lasso solution is also given by
\begin{align} \label{eqn:fused_as_gen}
    \Vhat{\beta} = \argmin{\vbeta \in \Real^P} 
        \frac{1}{2} ||\y - \X \vbeta||^2_2 + 
        \lambda_1 ||\D_{\lambda_2} \vbeta||
\end{align}
with the matrix $\D_{\lambda_2}$ having the form given in  \eqref{eqn:fused_D}. 

\subsection{Clustered Lasso}

While the fused lasso works for ordered coefficients, it is too restrictive for
most applications. This issue is addressed by the clustered lasso, which is
again a special case of the generalized lasso \citep{she2010sparse} .  The
clustered lasso estimates are given by:
\begin{align} \label{eqn:clustered_lasso}
    \Vhat{\beta} = \argmin{\vbeta \in \Real^P} 
        \frac{1}{2} ||\y - \X \vbeta|| + 
        \lambda_1 ||\vbeta||_1 + 
        \lambda_2 \sum_{i < j} |\beta_i - \beta_j|
\end{align}

While the clustered lasso is a simple combination of the convex clustering
penalty with a variable selection penalty, \citet{she2010sparse} shows that it
is inconsistent for identifying the correct clustering of $\vbeta$.  This issue
is due to the use of an $\ell_1$ relaxation, and \citet{she2010sparse} proposes
a multi-step algorithm, combining data augmentation and adaptive weights, to
remedy the issue \citep{zou2006adaptive}.  Given $\Vhat{\beta}_{a}$ and
$\Vhat{\beta}_{w}$ the new solution is found by solving the following
optimization problem:
\begin{align} \label{eqn:daw_lasso}
    \Vhat{\beta} = \argmin{\vbeta \in \Real^P}
        \frac{1}{2} ||\y^* - \X^* \vbeta||^2_2 + 
        \lambda_1 \sum w_i | (\D_{\lambda_2} \vbeta)_i|
\end{align}

where $\y^* = [\y\Tra, \y_a\Tra]\Tra$, and $\X^* = [\X \Tra, \X_a\Tra]\Tra$,
$\X_a = \sqrt{\tau} \mlambda \left( \Vhat{\beta}_{a} \right)$, $\y_a = \V{0}$,
$w_i = 1/|(\D_{\lambda_2} \Vhat{\beta}_w)_i|$, and $\mlambda(\vbeta)$ is given
by
\begin{align}
    \mlambda(\vbeta) = 
    \begin{cases}
        \I - \vbeta \vbeta\Tra / ||\vbeta||^2_2 & \text{ if }  \vbeta \neq \V{0} \\
        \I & \text{ if } \vbeta = \V{0}
    \end{cases}
\end{align}

The problem in Equation \ref{eqn:daw_lasso} is called the data augmented
weighted clustered lasso (DAW-CLASSO). \citet{she2010sparse} proposes solving
the problem multiple times to improve parameter estimates as given in Algorithm
\ref{alg:daw_classo}.

\RestyleAlgo{boxruled}
\LinesNumbered
\begin{algorithm}[ht]
    \caption{DAW-CLASSO \citep{she2010sparse}} \label{alg:daw_classo}
    Solve Problem \ref{eqn:clustered_lasso}. Call the estimate 
        $\Vhat{\beta}_c$ \\
    \Repeat{desired}{
        Sort $\Vhat{\beta}_c$ in ascending order \\
        Fit the fused lasso \eqref{eqn:fused_lasso} according
            the order in Step 3. Label the estimate $\Vhat{\beta}_f$ \\
        Substitute $\Vhat{\beta}_{a}$ with $\Vhat{\beta}_c$ and 
            $\Vhat{\beta}_f$ for $\Vhat{\beta}_{w}$ \\
        Create $\y^{*}$ and $\X^{*}$ and solve \eqref{eqn:daw_lasso}.
        Overwrite $\Vhat{\beta}_c$ with this estimate
    }
\end{algorithm}

Obvious computational bottlenecks exist in this approach. First, we have to
solve multiple problems to get one solution to \eqref{eqn:daw_lasso} and
repeating these steps is inefficient. Second, the DAW-CLASSO has three tuning
parameters, $\tau$, $\lambda_1$, and $\lambda_2$, and a grid search with
cross-validation needs to be incorporated to find their optimal values.

\subsection{OSCAR}

Another two part penalty which encourages sparsity and clustering in $\vbeta$
is OSCAR \citep{bondell2008oscar}. They propose combining an $\ell_1$ penalty
with an $\ell_\infty$ penalty, which encourages the absolute value of
coefficient pairs to be equal. The solution to their problem is given by:
\begin{align} \label{eqn:oscar}
    \Vhat{\beta} = \argmin{\vbeta \in \Real^P} 
        \frac{1}{2} ||\y - \X \vbeta||^2_2 + 
            \lambda_1 ||\vbeta||_1 + 
            \lambda_2 \sum_{i < j} \text{max} 
                \left\{ |\beta_i|, |\beta_j| \right\}
\end{align}

While this problem is convex, the penalty on $|\vbeta|$ increases its
computational complexity relative to methods which penalize differences in
$\vbeta$.  In their paper, \citet{bondell2008oscar} solve the problem by
expanding $\beta_j$ into $\beta^+_j - \beta^-_j$ ($\beta_j^+, \beta_j^- \geq
0$) and introducing auxiliary variables for the pairwise maxima. They estimate
$\vbeta$ using a quadratic program with $(P^2 + 3P)/2$ parameters and $P^2 + P
+ 1$ linear constraints.  \citet{zhong2012efficient} solve the problem with an
efficient the accelerated projected gradient method but it is not implemented
in any statistical software. 

However, like with the clustered and fused lasso, OSCAR can be cast as a
generalized lasso. Using the identity $\text{max}\{|\beta_j|, |\beta_k|\} =
\frac{1}{2} \{ |\beta_k - \beta_j| + |\beta_j + \beta_k| \}$ the penalty in
\eqref{eqn:oscar} becomes $\lambda_1 ||\vbeta||_1 + \frac{\lambda_2}{2} \sum_{i
< j} |\beta_i - \beta_j| + \frac{\lambda_2}{2} |\beta_i + \beta_j|$. An
appropriate $\D_{\lambda_2}$ like in \eqref{eqn:fused_D} can be created for
this penalty, and the generalized lasso path algorithm can be used for fixed
$\lambda_2$. 

\subsection{PACS}

\citet{sharma2013pacs} propose a penalty called pairwise absolute clustering
and shrinkage (PACS) to address problems with OSCAR. Like the naive clustered
lasso in \eqref{eqn:clustered_lasso}, the authors note that OSCAR is not an
oracle procedure \citep{fan2001variable}.  As done by \citet{she2010sparse} for
the clustered lasso, they propose an adaptive weighting scheme and show that
PACS is a method with oracle properties. The PACS coefficient estimates are
given by:
\begin{align} \label{eqn:pacs}
    \Vhat{\beta} = \argmin{\vbeta \in \Real^P} 
        \frac{1}{2} ||\y - \X \vbeta|| + 
        \lambda \left \{ \sum_{j = 1}^P w_j |\beta_j| + 
        \sum_{j < k} w_{jk(-)} |\beta_j - \beta_k| + 
        \sum_{j < k} w_{jk(+)} |\beta_j + \beta_k| \right \}
\end{align}

The PACS penalty combines an $\ell_1$ norm to encourage sparsity in the
coefficients and penalizes their sum and differences to encourage sparsity in
their absolute differences. This effect of PACS is similar to that of OSCAR
and, using the identity $\text{max}\{|\beta_j|, |\beta_k|\} = \frac{1}{2} \{
    |\beta_k - \beta_j| + |\beta_j + \beta_k| \}$, it can be shown that OSCAR
is a special case of PACS. 

There are two differentiating factors between OSCAR and PACS. First, PACS
incorporates non-negative weights $(w_j, w_{jk(+)}, w_{jk(-)})$ which allow an
analyst to incorporate additional prior information into the model. The authors
discuss two general methods for choosing weights: data adaptive weights and
weights related to the correlation between predictors.  Additionally, PACS
enjoys computational advantages relative to the penalization approaches
mentioned earlier. Like the penalties above, PACS can be recast as a
generalized lasso for an appropriate structure matrix $\D$.  However, unlike
the fused lasso, clustered lasso, and OSCAR, the resulting penalty matrix will
only have one tuning parameter. This allows the full coefficient path to be
estimated by only one run of the algorithm developed by
\citet{tibshirani2011genlasso}. 

\subsection{Other Methods}

The above list of methods is by no means exhaustive. Many other penalized
regression approaches exist, but rely on the assumption that correlated columns
have similar effects.  For example, \citet{witten2014cluster} propose the
Cluster Elastic Net, where clusters in $\vbeta$ are encouraged to form for
highly correlated predictors, and groups for the columns of $\X$ are learned in
a data adaptive fashion. \cite{li2018graph} propose a three-part penalty to
estimate coefficients by assuming an underlying graph structure in $\X$ and use
an estimate of its covariance for adaptive weighting. Their method incorporates
many other approaches as special cases and we refer the reader to their paper
for details.

\subsection{Issues with penalization based approaches}

The methods discussed in this section suffer from three major issues. First,
regardless of the asymptotic proofs provided by \citet{she2010sparse} and
\citet{sharma2013pacs}, no method gives an exact clustering of the regression
coefficients in finite samples. To estimate a clustering, the above approaches
need to be augmented with a procedure such as $k$-means or hierarchical
clustering, but choosing the number of clusters remains an unresolved question.
Second, none of the methods mentioned above provide clear guidance on
uncertainty quantification for the coefficient estimates. The default choice
in most situations would be to use the bootstrap \citep{efron1994introduction}
but it is known to fail for $\ell_1$ penalties \citep{kyung2010}. Finally,
tuning parameter selection can be a computational bottleneck for the methods
discussed; all but PACS have a tuning parameter in the structure matrix, $\D$,
and finding their optimal values requires the use of a grid search. Resolution
of these issues from the frequentist perspective are open problems. Their
solutions have the potential for widespread use since frequentist methods tend
to be computationally efficient relative to their Bayesian counterparts. 

\subsection{Bayesian corollaries of penalized regression}

The issues of uncertainty quantification and tuning parameter selection can be
resolved in the Bayesian context, albeit at greater computational cost. All
methods discussed in this section have Bayesian analogs for which Gibbs
samplers can be used. Such methods would provide an estimate of the posterior
distribution for all parameters and uncertainty quantification would include
uncertainty regarding the tuning parameter. Additionally, these methods would
allow for the use of priors such as the Horseshoe
\citep{carvalho2010horseshoe}, Horseshoe+ \citep{bhadra2017horseshoe+}, and
Dirichlet-Laplace \citep{bhattacharya2015dirichlet}, all of which have superior
shrinkage properties relative to the Laplace distribution.

Unfortunately, even Bayesian penalized regression would fail to provide a
clustering of the data without using a second step. This issue can be remedied
by the use of mixture models as discussed in Section \ref{sec:mix_models}. 

\section{Mixture Models} \label{sec:mix_models}

The methods mentioned in the previous sections have been unsatisfactory in a
few ways. Two-step processes do not solve the coefficient clustering problem
directly while penalization based approaches fail to set coefficients exactly
equal to each other in finite samples. Additionally, in the frequentist setup,
uncertainty quantification remains an open problem for both approaches. These
issues are best resolved in the Bayesian context with the use of mixture
distributions as a prior for $\vbeta$. 

In this section we first discuss finite mixture models and how they can be used
to find clusters in $\vbeta$. We then extend this discussion to countably
infinite mixture models and the use of Dirichlet Process priors for the
regression coefficients. 

\subsection{Finite Mixture Models}

Finite mixture models provide a useful framework for clustering a vector of
points $\{y_1, \dots, y_N \}$ \citep{murphy2012machine}. Assuming we want to
partition the data into $K$ underlying groups, we can model the $y_i$ as a
mixture distribution of the form $f(y_i | K, \vpi,\vtheta) = \sum_{k = 1}^K
\pi_k f_k (y_i | \theta_k)$ where $\vtheta = \{ \theta_1, \dots, \theta_k \}$
is a set of parameters which index the $k$'th base distribution, $f_k$, and
$\vpi$ is a vector of mixing weights with $\pi_i \geq 0$ and $\sum_i \pi_i =
1$. 

Finite mixture models are commonly given a latent class representation where
the elements of a vector, $\c$, are used to encode class membership for the
elements of $\y$. Because our end goal is coefficient clustering in normal
linear models, assume that each base distribution is normal with varying mean
and constant variance. The finite mixture model can be then be written
hierarchically as follows: 
\begin{align} 
\label{eqn:fmm_hier}
\begin{split}
    y_i | c_i = k, \vtheta & \sim \normal(\theta_k, \sigma^2) \\
    c_i | \vpi & \sim \text{Discrete}(\pi_1, \dots, \pi_K) \\
    \vpi | \alpha, K & \sim 
        \text{Dirichlet} \left( \frac{\alpha}{K} \V{1}_K \right) \\
    \theta_k & \sim G_0
\end{split}
\end{align}

From the above it can be seen that the mean of each observation, $\mu_i$, is
equal to $ \sum_{k = 1}^K I(c_i = k) \theta_k$, and that the elements of the
mean vector, $\vmu = (\mu_1, \dots, \mu_N)\Tra$, can take on only one of $K$
values. Therefore, we can view $\c$ as a vector of class labels for the
elements of $\vmu$ instead of $\y$. Additionally, because $\y | \vmu, \sigma^2
\sim \normal(\I_N \vmu, \sigma^2 \I_N)$, we can easily modify the hierarchy in
\eqref{eqn:fmm_hier} to incorporate a covariate matrix $\X$ and parameter
vector $\vbeta$, as follows:
\begin{align}
\label{eqn:nott_hier}
\begin{split}
    \y | \vbeta, \X, \sigma^2 & \sim \normal(\X \vbeta, \sigma^2 \I) \\
    (\forall p \in \{1, \dots, P \}) \ \beta_p | c_p, \vtheta & =
        \sum_{k = 1}^K I(c_p = k) \theta_k \\
    (\forall p \in \{1, \dots, P \}) \ c_p | \vpi & \sim 
        \text{ Discrete}(\pi_1, \dots, \pi_K) \\
    \vpi | \alpha & \sim \text{Dirchlet} 
        \left(\frac{\alpha}{K} \V{1}_K \right) \\
    (\forall k \in \{1, \dots, K \}) \ \theta_k | \gamma & \sim
        \normal(0, \gamma^2)
\end{split}
\end{align}
Here $\c$ is a $P \times 1$ vector which stores class labels for the elements
of $\vbeta$. 

The hierarchy in \eqref{eqn:nott_hier} further implies that $\beta_p \sim G =
\sum_{k = 1}^K \pi_k \delta_{\theta_k}(\beta_p)$ where $\delta_{\theta_k}(x)$
is 1 if $x = \theta_k$ and $0$ otherwise. Consequently, each element of
$\vbeta$ is a finite mixture of K delta functions, and can only take on one of
$K$ values.  If we know $K$ in advance, estimating $\vtheta$ and class
memberships, $\c$, provides a solution to the coefficient clustering problem. 

Unfortunately, in almost all applications $K$ is unknown and we need to
estimate the number of clusters in $\vbeta$.  One way to do this is to fit the
model in \eqref{eqn:nott_hier} for different values of $K$ and use a model
selection criteria such as cross validation. However, since we are focusing our
attention to the Bayesian context, we can put a prior on the distribution $G$,
called a Dirichlet Process, which returns discrete distributions as its
samples. The Dirichlet Process is indexed by a concentration parameter,
$\alpha$, and a base measure, $G_0$, and its use in the coefficient clustering
problem will be our focus for the next three sections. 

\subsection{Dirichlet Processes}

The Dirichlet Process (DP) was first investigated by \citet{ferguson1973} and
\citet{antoniak1974}.  Using a stick-breaking construction of the DP, it can be
shown that the distribution $G(x) = \sum_{k = 1}^\infty \pi_k
\delta_{\theta_k}(x)$ is a sample from a $DP(\alpha, G_0)$, where $\pi_k = V_k
\prod_{l = 1}^{k - 1} (1 - V_l)$, $V_k \simiid \text{Beta}(1, \alpha)$, and
$\theta_k \simiid G_0$ \citep{sethuraman1994constructive}.  Hence,
distributions sampled from the Dirichlet Process are a countably infinite
mixture of point masses, which is a generalization of the finite mixture prior
for $\vbeta$ discussed above.

If we use the Dirichlet Process to allow the finite mixture model in
\eqref{eqn:fmm_hier} to have a variable number of clusters, the resulting model
is called a Dirichlet Process mixture model. Computationally efficient methods
for estimating such models were developed by \citet{escobar1995bayesian},
\citet{maceachern1998estimating}, and \citet{neal2000markov}.  The model
hierarchy is given below:
\begin{align}
\label{eqn:dpmm_hier}
\begin{split}
    y_i | \mu_i & \sim  f(\mu_i) \\
    \mu_i | G & \sim G \\
    G & \sim DP(G_0, \alpha)
\end{split}
\end{align}

A key insight into the behaviour of the DP was given by
\citet{blackwell1973ferguson} who show that, in exchangeable models, the prior
distribution of $\mu_i$ conditional on all other observations is given by
\begin{align} 
\label{eqn:dir_prior_black}
    \mu_i| \vmu_{-i} & \sim 
        \frac{1}{N - 1 + \alpha} 
        \sum_{i \neq j} \delta_{\mu_j}(\mu_i) + 
        \frac{\alpha}{N - 1 + \alpha} G_0
\end{align}
where $\vmu_{-i}$ is the vector $\vmu$ without the element $\mu_i$.
\cite{neal2000markov} gives a similar conditional prior for the class labels,
$\c$, by viewing the Dirichlet process mixture model as the limit of finite
mixture models as $K \to \infty$.  By integrating out the mixing parameters
$\vpi$ he derives the following conditional priors for the elements of $\c$
\begin{align}
\label{eqn:neal_prior}
\begin{split}
    P(c_i = k | \c_{-i} ) = \frac{n_{-i, k}}{N - 1 + \alpha} \\
    P(c_i \neq c_j \text{ for all } i \neq j | c_{-i}) = 
        \frac{\alpha}{N - 1 + \alpha}
\end{split}
\end{align}
where $n_{-i, k}$ is the number of $c_j$, $j \neq i$, that are equal to $k$.
These representations allow us to bypass the sampling of mixing weights,
$\vpi$, and show that the prior conditional probability of an observation being
assigned to an existing cluster is proportional to the number of elements in
the cluster. 

\subsection{Dirichlet Process Priors for Regression} \label{sec:nott_dir}

The hierarchy in \eqref{eqn:nott_hier} can be modified to include a Dirichlet
Process prior for $\vbeta$. \citet{nott2008predictive} explored the estimation
and predictive capacity of the DP prior with a normal base measure along with
its applications in penalized spline smoothing.  The full model hierarchy of
his paper is as follows:
\begin{align}
\label{eqn:lm_dp}
\begin{split}
    \y | \vbeta, \X, \tau & \sim \normal(\X \vbeta, \tau\Inv \I) \\
    (\forall p \in \{1, \dots, P \}) \ \beta_p | G & \sim G \\
    G & \sim DP(G_0, \alpha) \\
    G_0 | \lambda & = \normal(0, \lambda\Inv) \\
    \tau & \sim \mathcal{G}(a_{\tau}, b_{\tau}) \\
    \lambda & \sim \mathcal{G}(a_{\lambda}, b_{\lambda}) \\
    \alpha & \sim \mathcal{G}(a_{\alpha}, b_{\alpha})
\end{split}
\end{align}

\citet{nott2008predictive} derived an efficient Gibbs sampler for the model in
\eqref{eqn:lm_dp} by leveraging a latent class representation for group
membership. His algorithm iterates between sampling the cluster indicators for
the elements of $\vbeta$, $\c$, and all other model parameters conditioned on
the number of clusters in $\c$, $K$. We discuss the salient details of the
algorithm below and refer the reader to the original paper for a full
derivation. 

Conditional on $\c$ we need to estimate only $K$ true effects, $\vtheta$. We
can define a $P \times K$ matrix $\C$ whose rows are one hot encoded with class
membership labels ($\C_{pk} = 1$ if $c_p = k$ and 0 otherwise) and see that $\y
\sim \normal(\X \C \vtheta, \tau\Inv \I_N)$ with $\vbeta = \C \vtheta$. Since
the elements of $\vtheta$ are samples from $G_0$, it has a $K$ dimensional
multivariate normal prior and its update is a simple Bayesian linear regression
update with a normal prior.

Updating $\c$ is the most computationally intensive step of the Gibbs sampler
because its elements have to be updated sequentially.  To update the class
label for a predictor, $c_p$, we need to calculate posterior probabilities of
the predictor belonging to each cluster. Here, the number of clusters is $K +
1$ where $K$ is the number of unique elements in $\c_{-p}$.
\citet{nott2008predictive} updates $c_p$ by integrating out the parameter
vector $\vtheta$, which is important for the mixing of the Markov chains, but
requires $K + 1$ matrix inversions for each predictor. The probability that
$c_p = k$ is given by: 
\begin{align}
\label{eqn:c_up}
\begin{split}
    P(c_p = k | \c_{-i}, \tau, \lambda, \alpha, \vtheta, \X, \y) & 
        \propto f(\y | \tau, \lambda, \c', \X)
        P(c_p = k | \c_{-i}, \alpha) \\
    f(\y | \tau, \lambda, \c', \X) & = \int f(\y | \tau, \vtheta, \X, \C', \tilde{K})
        f(\vtheta | \lambda, \tilde{K}) d \vtheta \\ 
    & \propto \int \tau^{N/2} 
        \exp \left \{ -\frac{\tau}{2} ||\y - \X \C' \vtheta||^2_2 \right \} 
        (\lambda)^{\tilde{K} / 2} \exp \left\{-\frac{\lambda}{2} ||\vtheta||^2_2 \right\}
        \d \vtheta \\
    & = \lambda^{N/2} \left( \frac{\lambda}{\tau} \right)^{\tilde{K}/2} 
        |\A|^{-1/2} 
        \exp \left\{ -\frac{\tau}{2} 
        \left (\y\Tra \y - \y\Tra \Xtilde \A\Inv \Xtilde\Tra \y \right)
        \right \}
\end{split}
\end{align}
where $\A = \Xtilde\Tra \Xtilde + \frac{\lambda}{\tau} \I$, $\Xtilde = \X \C'$,
$\c'$ is the vector of class labels with $c_p = k$, $\tilde{K}$ is the number
of clusters in $\c'$, $\vtheta$ is a $\tilde{K} \times 1$ vector, and $P(c_p =
k | \c_{-i}, \alpha)$ can be calculated by using the formulas in
\eqref{eqn:neal_prior}. It should be noted that while $K + 1$ inverses have to
be computed for each predictor, they can be done in parallel to speed up the
algorithm.

\subsection{Point mass + Dirichlet Process Prior}

All the penalization methods discussed in Section \ref{sec:pen_methods} consist
of penalties which aim to perform variable selection and clustering
simultaneously. \citet{dunson2008bayesian} propose a prior for the same effect,
using a mixture of a point mass at zero with a distribution given a DP prior
\begin{align} \label{eqn:dunson_prior}
    G = \pi_0 \delta_0 + (1 - \pi_0) G^*, \ G^* \sim DP(\alpha, G_0)
\end{align}
where $0 \leq \pi_0 \leq 1$ is a mixture weight, $\delta_0$ is a point mass
probability density at $0$, and $DP(\alpha, G_0)$ is a Dirichlet Process prior
with precision $\alpha$ and base measure $G_0$. Using the constructive
definition of \citet{sethuraman1994constructive} the prior in
\eqref{eqn:dunson_prior} is again an infinite mixture of point masses with the
first point mass fixed at zero
\begin{align} \label{eqn:dunson_inf_pm}
    G = \pi_0 \delta_0 + (1 - \pi_0) \sum_{k = 1}^{\infty} \pi_k
    \delta_{\theta_k} = \sum_{k = 0}^{\infty} \tilde{\pi}_k \delta_{\theta_k}
\end{align}
where $\tilde{\pi}_0 = \pi_0$, $\tilde{\pi}_k = (1 - \pi_0) \pi_k$ if $k \geq
1$, $\theta_0 = 0$ and $\theta_k \simiid G_0$ if $k \geq 1$. 

Computation for this model is similar to the model discussed in Section
\ref{sec:nott_dir} because the prior in \eqref{eqn:dunson_inf_pm} is a
clustering prior with the restriction that the first cluster is zero. We can
account for this restriction by making a few changes to the Gibbs sampler
developed by \citet{nott2008predictive}. To update $\c$, conditional prior
probabilities of the form in \eqref{eqn:neal_prior} need to be determined and
the integral in \eqref{eqn:c_up} has to be modified by dropping, from $\X$, the
columns in the null (zero) cluster. The update of $\vtheta$ also relies on
dropping the columns in the null cluster because their effects are restricted
to be zero; the rest of the update proceeds as before with the smaller
predictor matrix. 

If we let $P_0$ denote the number of variables which are in the null cluster
and $P_c$ denote the number of variables which are non-zero $(P = P_0 + P_c)$,
we can calculate conditional prior probabilities of the form in
\eqref{eqn:neal_prior} by thinking of the prior in two levels: the first being
a two component finite mixture model and the second a Dirichlet Process. To
begin, we need to determine the prior probabilities of the predictor $p$ being
equal to zero, $P(c_p = 0 | \c_{-p}, \alpha_0)$, where $[\pi_0, (1 - \pi_0)]$
is given a $ \text{Dirichlet} \left( \frac{\alpha_0}{2}, \frac{\alpha_0}{2}
\right)$ prior, with $\pi_0 \sim \text{Unif}(0, 1)$ if $\alpha_0 = 2$. This is
given directly by \citet{neal2000markov} to be 
\begin{align} \label{eqn:fm_dp}
    \pi^*_0 = \pi(c_p = 0 | \c_{-p}, \alpha_0) & = 
        \frac{n_{-p, 0} + \alpha_0 / 2}{P - 1 + \alpha_0}
\end{align}
Because only a subset of the predictors are non zero, the Dirichlet Process
prior can be thought of as active for only $P_c$ predictors. Therefore, the
probabilities from \eqref{eqn:fm_dp} can be combined with the probabilities
from \eqref{eqn:neal_prior} to get the conditional prior for cluster membership
\begin{align}
\begin{split}
    \pi^*_0 = P(c_p = 0 | \c_{-p}, \alpha_0) & = 
        \frac{n_{-p, 0} + \alpha_0 / 2}{P - 1 + \alpha_0} \\
    \text {For $k \geq 1$, } P(c_p = k | \c_{-p}, \alpha) & = 
        (1 - \pi_0^*) \left( \frac{n_{-p, k}}{P_c - 1 + \alpha} \right) \\
    P(c_p \neq 0 \text{ and } c_p \neq c_j \text{ for all } p 
        \neq j| \c_{-p}, \alpha) & = 
        (1 - \pi_0^*) \left( \frac{\alpha}{P_c - 1 + \alpha} \right)
\end{split}
\end{align}

\subsection{Issues with Mixture Model Approaches}

The primary issue with the approaches in this section is their computational
complexity.  Using Bayesian methods with mixture models as priors for $\vbeta$
allows the simultaneous estimation of class membership and uncertainty, but it
comes at a cost.  While the inversion of multiple matrices in each iteration of
the class label update is parallelizable, updating the cluster means,
$\vtheta$, is not. Therefore, there will always be a step of the Gibbs sampler
with computational complexity cubic in the number of clusters. It is reasonable
to expect that the number of clusters in $\vbeta$ will increase with its size,
making such methods computationally infeasible for even moderate $P$. Possible
remedies for this situation could be the use of posterior approximation
algorithms such as Variational Bayes \citep{blei2017variational} or Approximate
Bayesian Computation \citep{beaumont2010approximate}, but their development in
the context of Dirichlet Process priors for regression remains an open problem. 

\section{Applications}

The methods above have primarily been used to resolve issues with
multicollinearity in the predictors. Most early applications were with genetic
data, which motivated the development of two-step and most penalized regression
approaches. For example, \citet{hastie2001harvest} apply a two-step approach to
a lymphoma dataset with approximately 3000 gene expression and 36 patients
while \citet{zou2005enet} apply the elastic net to a leukemia dataset with
approximately 7000 genes and 70 patients. Another class of problems which can
be addressed with such methods arise in epidemiological research, where a large
number of correlated exposures are related to a response. For example,
\citet{maclehose2007bayesian} apply a Dirichlet Process prior with a point mass
at zero to a dataset where the effect of exposure to 18 herbicides on retinal
degeneration in approximately 30,000 women was of interest. 

A more recent application of these methods has been in the analysis of fMRI
data. Technological advances have allowed the collection of large datasets
where the columns of $\X$ consist of measurements from highly correlated
sections of the brain, called voxels.  \citet{li2018graph} apply their method
to an fMRI study where subjects were shown pictures with or without the
presence of a human face.  Blood-oxygen-level dependent imaging was
concurrently conducted to measure activation of neurons, with the goal of
classifying the presence of a face in the picture.  The final analysis was
conducted separately for each individual, where $N = 90$ and $P$ was greater
than 5000 for each patient. 

\section{Conclusion}

This paper provides an introduction to many modern dimension reduction
techniques which aim to simultaneously select relevant predictors and find
groups with the same effect on the response.  Such methods were initially
developed to address problems which arose in genetic data analysis, but have
since been extended to epidemiology and fMRI research.  They effectively
overcome limitations of popular methods such as the lasso, which often fail to
select all predictors related to a response.  

In addition to an overview of the methods, a discussion of their respective
limitations was also provided. Two-step approaches, while a logical first step
to the clustering problem, fail to directly cluster the coefficient vector.
Penalized regression approaches, while computationally efficient, are not
adequate when an exact clustering of the predictors or uncertainty
quantification is required. Finally, while these issues can be effectively
resolved in a Bayesian context via the use of Dirichlet Process priors, their
use is computationally infeasible for many modern problems, such as the fMRI
dataset analyzed by \citet{li2018graph}. 

Ideally, future research will focus on algorithmic developments which allow the
application of such models to large datasets, while incorporating parmeter
estimation, uncertainty quantification, and parameter clustering into one
procedure. A fruitful avenue of research could be the development of
Variational Bayes algorithms or the exploitation of modern computing
technologies, such as GPUs. 

\bibliography{references} 
\bibliographystyle{plainnat}

\end{document}